\documentclass[apj]{emulateapj}

\slugcomment{}

\shorttitle{}
\shortauthors{}

\usepackage{color}
\usepackage{amsmath}
\usepackage{booktabs}

\begin{document}

%% LaTeX will automatically break titles if they run longer than
%% one line. However, you may use \\ to force a line break if
%% you desire.

\title{Widening of Protostellar Outflows: an Infrared Outflow Survey in Low Luminosity Objects}

\author{Tien-Hao Hsieh$^{1,2}$, Shih-Ping Lai$^{1,2}$, and Arnaud Belloche$^{3}$}
\affil{$^{1}$Institute of Astronomy, National Tsing Hua University (NTHU), Hsinchu 30013, Taiwan}
\affil{$^{2}$Institute of Astronomy and Astrophysics, Academia Sinica, P.O. Box 23-141, Taipei 106, Taiwan}
\affil{$^{3}$Max-Planck-Institut f\"{u}r Radioastronomie, Auf dem H\"{u}gel 69, 53121 Bonn, Germany}

\email{thhsieh@asiaa.sinica.edu.tw; slai@phys.nthu.edu.tw}

\begin{abstract}
%Aims
We present an outflow survey toward 20 Low Luminosity Objects (LLOs), namely protostars with an internal luminosity lower than 0.2 $L_\odot$.
Although a number of studies have reported the properties of individual LLOs, the reasons for their low luminosity remain uncertain.
To answer this question, we need to know the evolutionary status of LLOs.
Protostellar outflows are found to widen as their parent cores evolve, and therefore, the outflow opening angle could be used as an evolutionary indicator.
%Methods & Results
The infrared scattered light escapes out through the outflow cavity and highlights the cavity wall, giving us the opportunity to measure the outflow opening angle.
Using the Canada-France-Hawaii Telescope, we detected outflows toward eight LLOs out of 20 at Ks band, and 
based on archival \textit{Spitzer} IRAC1 images, we added four outflow-driving sources from the remaining 12 sources.
By fitting these images with radiative transfer models, we derive the outflow opening angles and inclination angles.
To study the widening of outflow cavities, we compare our sample with the young stellar objects from \citet{ar06} and \citet{ve14} in the plot of opening angle versus bolometric temperature taken as an evolutionary indicator.
%Conclusions
Our LLO targets match well the trend of increasing opening angle with bolometric temperature reported by Arce \& Sargent and are broadly consistent with that reported by Velusamy et al., suggesting that the opening angle could be a good evolutionary indicator for LLOs.
Accordingly, we conclude that at least 40\% of the outflow-driving LLOs in our sample are young Class 0 objects.
\end{abstract}

%% Keywords should appear after the \end{abstract} command. The uncommented
%% example has been keyed in ApJ style. See the instructions to authors
%% for the journal to which you are submitting your paper to determine
%% what keyword punctuation is appropriate.


\section{INTRODUCTION}
%outflow widen 
%Protostellar outflows are believed to be largely responsible for the removal of angular momentum from the collapsing cores, thus they play an important role in the star formation process.
Protostellar outflows are commonly seen toward Young Stellar Objects (YSOs), especially at the embedded stage. 
Studying the outflow properties allows us to probe the nature of the 
central protostars indirectly. 
In a protostellar core, the outflow can carve out a biconical cavity that widens as the core evolves \citep[hereafter AS06 and VLT14 for the first and last references, respectively]{ar06,of11,ve14}.
The widening may initially originate from the precession of a collimated jet, but the small precession angles found in protostars cannot account for the large opening angles observed at a more evolved stage \citep{re00,ar04}. 
\citet{ar04}, therefore, suggest that the widening of the outflow cavity is likely produced by the widening of the stellar wind from the central YSO.

%Infrared scattered light
The near-infrared (NIR) continuum emission traces the scattered light that escapes through the outflow cavity, and has thus offered us the opportunity to study the outflow structures over the past decades \citep{lu97,pa99,ei05,te06,st06,to07,se08,ve14}.
For a given outflow opening angle and a given inclination angle, the scattered light image can be modeled with radiative transfer codes \citep{wh03a,wh03b,ro07}.
Therefore, comparing the synthetic images with observed NIR images allows us to derive the outflow opening angles, and thus study their evolution.

%VeLLOs
Very Low Luminosity Objects (VeLLOs), first discovered by the \textit{Spitzer} Space Telescope \citep{yo04}, are defined as embedded protostars with an internal luminosity $L_{\rm int}$ $<$ 0.1 $L_{\odot}$ \citep{di07}; the internal luminosity is the total luminosity of the central protostar and circumstellar disk and, at the early stage, is likely dominated by the photospheric and accretion luminosities.
Using the data from the \textit{Spitzer} Legacy Project ``From Molecular Cores to Planet Forming Disks'' (c2d; Evans et al.\ 2003, 2009), \citet{du08} identified 15 VeLLOs in the Perseus, Ophiuchus, Serpens, Lupus, and Chamaeleon molecular clouds plus 82 regions that contain 95 small, dense cores.
During the last decade, studies of individual VeLLOs found
that their properties vary much from one object to the other
(IRAM 04191: Andr\'{e} et al.\ 1999; Belloche et al. 2002; Dunham et al.\ 2006,
L1014: Bourke et al.\ 2005; Huard et al.\ 2006,
L1521F: Bourke et al.\ 2006; Takahashi et al.\ 2013,
Cha-MMS1: Belloche et al.\ 2006, Tsitali et al.\ 2013,
L328: Lee et al.\ 2009, 2013,
L673-7: Dunham et al.\ 2010a,
CB130: Kim et al.\ 2011,
L1148: Kauffmann et al.\ 2011,
IC 348-SMM2E: Palau et al.\ 2014,
IRAS\,16253: Hsieh et al.\ 2016).
Based on these studies, \citet{du14} summarize three interpretations of the low luminosity, 
suggesting that VeLLOs can be (1) very low mass protostars, (2) extremely young protostars, or (3) protostars in a quiescent phase of the episodic accretion process, in which a protostar is at a quiescent accretion phase for most of the time and accretion bursts occasionally occur to deliver material onto the central protostar \citep{ke95,le07,du10b,du12,jo15,ki16}.
\citet{sc12} conducted an outflow survey of VeLLOs, and from the outflow forces, derived the time-averaged accretion luminosities;
they found that these time-averaged accretion luminosities are higher than their current internal luminosities, suggesting that VeLLOs are in a quiescent phase of an episodic accretion process.
On the other hand, using the N$_2$D$^+$/N$_2$H$^+$ abundance ratio as a chemical evolutionary indicator and the line width as a dynamical evolutionary indicator, \citet{hs15} suggest that VeLLOs tend to be young Class 0 protostars.

In this paper, we aim at studying the evolution of the outflow opening angle in protostellar objects at early stage.
We describe the sample and the observations in Section \ref{sec:obs}.
In Section \ref{sec:result}, we report the observational results and describe how we derive the outflow opening angles using the radiative models from \citet{wh03a,wh03b}.
In Section \ref{sec:discussion}, we discuss the correlation of the derived outflow opening angle with bolometric temperature ($T_{\rm bol}$, in comparison with AS06 and VLT14) and with the ratio of bolometric to submillimeter luminosity ($L_{\rm bol}/L_{\rm smm}$), both taken as evolutionary indicators.
Finally, we summarize these results in Section \ref{sec:summary}.

\tabletypesize{\scriptsize}
\tabcolsep=0.08cm
\begin{deluxetable*}{cccccccccccc}
\tabletypesize{\tiny}
\tablecaption{Source properties}
\tablehead{ 
\colhead{Source}	& \colhead{Other name}	&	\colhead{R.A.}	& \colhead{Dec}	& \colhead{$L_\textmd{int}$}	& \colhead{$T_\textmd{bol}$} & \colhead{d}	& \colhead{Ks-band}	& \colhead{H$_2$ jet} & \colhead{Op. angle} & \colhead{Inc. angle}	& Ref.\\
\colhead{}		&		\colhead{}		& \colhead{}	& \colhead{}	& \colhead{(L$_{\odot}$)}		& \colhead{(K)}	& \colhead{pc}	& \colhead{image}	&	\colhead{}	& \colhead{($^\circ$)}		& \colhead{$^\circ$)}	& \colhead{}
}
\startdata 
DCE 001	& IRAM 04191	& 04:21:56.88 & +15:29:46.0 & 0.05	      & 28 (1) 		& 140		& U$^*$	& Y		& 25$^{+7.5}_{-7.5}$	& 20$^{+2.5}_{-17.5}$		& 1, 2	\\
DCE 004	& L1521F		& 04:28:38.90 & +26:51:35.6 & 0.03		& 20 (1)		& 140		& U$^*$	& N	& 60$^{+2.5}_{-7.5}$	& 15$^{+2.5}_{-2.5}$		& 4, 11\\
DCE 024	& CB130-1-IRS1	& 18:16:16.39 & -02:32:37.7 & 0.07	& 56 (7) 		& 270		& O		& N	& 15$^{+2.5}_{-2.5}$	& 55$^{+2.5}_{-2.5}$		& 9\\      
DCE 025	& L328-IRS	& 18:16:59.47 & -18:02:30.5 & 0.07		& 68 (7)		& 270 		& O		& N	& 25$^{+2.5}_{-2.5}$	& 50$^{+2.5}_{-2.5}$		& 6, 10\\
DCE 031	& L673-7		& 19:21:34.82 & +11:21:23.4 & 0.04		& 24 (4)		& 300		& U		& N	& -				& -		& 7\\
DCE 032	& L1148-IRS	& 20:40:56.66 & +67:23:04.9 & 0.09		& 145 (6)		& 325		& E			& N	& -				& -		& 8	\\
DCE 038	& L1014-IRS	& 21:24:07.60 & +49:59:08.9 & 0.09		& 67 (14)		& 250		& O		& N		& 65$^{+2.5}_{-2.5}$	& 20$^{+2.5}_{-2.5}$	& 3, 5\\
DCE 063	&				& 03:27:38.26 & +30:13:58.8 & 0.2		& 199 (26)		& 250		& E			& N	& -				& -	&		\\
DCE 064	& 				& 03:28:32.57 & +31:11:05.3 & 0.03		& 65 (7)		& 250		& O		& N	& 55$^{+2.5}_{-2.5}$	& 20$^{+2.5}_{-2.5}$	&	\\
DCE 065	& 				& 03:28:39.10 & +31:06:01.8 & 0.02		& 29 (1)		& 250		& U		& N	& -				& -			&\\ 
DCE 078	&				& 03:29:23.47 & +31:33:29.5 & 0.2		& 60 (5)		& 250		& O		& N	& 40$^{+2.5}_{-2.5}$	& 20$^{+2.5}_{-2.5}$	&\\
DCE 081	&				& 03:30:32.69 & +30:26:26.5 & 0.06		& 33 (2)		& 250		& U		& N	& -				& -			& \\      
DCE 090	&				& 03:32:29.18 & +31:02:40.9 & 0.2		& 114 (13)		& 250		& O		& N		& 45$^{+2.5}_{-2.5}$	& 25$^{+2.5}_{-2.5}$ &		\\      
DCE 092	&				& 03:33:14.38 & +31:07:10.9 & 0.14		& 47 (5)		& 250		& U$^*$	& N	& 25$^{+12.5}_{-2.5}$	& 15$^{+7.5}_{-12.5}$		&	\\
DCE 107	&				& 03:44:02.40 & +32:02:04.9 & 0.15		& 77 (3)		& 250 		& O		& Y		& 55$^{+2.5}_{-27.5}$& 5$^{+2.5}_{-2.5}$			& \\
DCE 109	&				& 03:44:21.36 & +31:59:32.6 & 0.11		& 348 (14)		& 250		& E			& Y		& -				& -		&	\\
DCE 145	& 				& 15:40:51.62 & -34:21:04.7  & 0.03		& \multicolumn{7}{l}{DCE 145 is identified as a galaxy based on the Ks-band image}		\\
DCE 181	& 				& 16:26:48.48 & -24:28:38.6 & 0.05		& 429 (23)		& 125		& P			& N	& -				& -		&	\\
DCE 182	& LFAM 26	& 16:27:05.23 & -24:36:29.5 & 0.15		& 105 (6)		& 125		& O		& N		& 70$^{+2.5}_{-2.5}$	& 5$^{+2.5}_{-2.5}$			& \\      
DCE 185	& IRAS 16253-2429	& 16:28:21.60 & -24:36:23.4 & 0.09& 31 (1) 		& 125		& U$^*$	& Y		& 35$^{+2.5}_{-2.5}$	& 5$^{+2.5}_{-2.5}$			&
\enddata
\tablecomments{
Col. (8): Outcome of the Ks-band observation: ``U'' stands for undetected source, ``O'' for outflow cavity detected, ``P'' for point source, and ``E'' for extended source. 
Col. (9): ``Y'' stands for  H$_2$ jet detection and ``N'' for no jet detected. 
Col. (10): Outflow opening angle derived from fitting with Whitney's model.
Col. (11): Best-fit inclination angle of the outflow axis with respect to the plane of the sky.
References:
(1) \citet{an99};
(2) \citet{be02};
(3) \citet{bo05};
(4) \citet{bo06};
(5) \citet{hu06};
(6) \citet{le09};
(7) \citet{du10a};
(8) \citet{ka11};
(9) \citet{ki11};
(10) \citet{le13};
(11) \citet{ta13}
}
\tablenotetext{*}{The outflows are detected with \textit{Spitzer} but not with CFHT.}
%\tablenotetext{a}{}
%\tablenotetext{b}{}
\label{tab:targets}
\end{deluxetable*}

%H2 jet
\section{OBSERVATIONS}
\label{sec:obs}
\subsection{Sample}
We selected the faintest targets from the catalog in \citet{du08} which includes 50 low-luminosity protostars with $L_{\rm int}\leq1.0\ L_\odot$ (15 have $L_{\rm int}\leq0.1\ L_\odot$) in five nearby molecular clouds ($d<400$\,pc) mapped by the c2d team \citep{ev03,ev09}.
Hereafter we define the source name as the initials of the first three authors followed by the source number in \citet{du08}, e.g., DCE 185.
Our sample includes 13 VeLLOs out of 15 from \citet{du08}, in which DCE 018 and 161 are excluded due to their low elevation ($<$35\arcdeg) from the Canada-France-Hawaii Telescope (CFHT) on Mauna Kea, Hawaii.
In addition to these 13 VeLLOs, we select 7 targets out of 9 in Dunham's catalog with $0.1\ L_{\odot} \leqslant L_{\rm int} \leqslant 0.2\ L_{\odot}$, namely Low Luminosity Objects (hereafter LLOs) in order to enlarge our sample.
Table \ref{tab:targets} lists the internal luminosities and the bolometric temperatures ($T_{\rm bol}$) that separate the targets into 13 Class 0 objects ($T_{\rm bol}<$ 70 K) and 7 Class I objects (70 K $\leqslant T_{\rm bol}\leqslant$ 650 K).
We note that this sample includes all 15 objects of our previous study of the envelope properties \citep{hs15}.

\subsection{Ks-band images}
\label{sec:obsk}
The observations were carried out using the Wide-field InfraRed Camera (WIRCam) at the 3.6 m CFHT in the Ks broad-band filter in March and April 2010.
The WIRCam has a field of view (FOV) of $20\arcmin\times20\arcmin$ with a sampling of 0\farcs3 per pixel.
The Ks-band filter has a bandwidth of 0.325 $\mu$m and a central wavelength of 2.146 $\mu$m.
Every target was observed with a cycled dither pattern of five positions. 
The exposure time per image was 25 s, and the total exposure time was 300--900\,s per source depending on its distance.
Saturated sources were later observed with several exposures of 5\,s in July and October 2011.
The seeing during our observations was about 0\farcs7 from 0\farcs5 to 1\farcs2.
The data were processed with the CFHT WIRCam standard pipeline.
The data reduction was done with the TERAPIX\footnote{http://terapix.iap.fr} software, that corrects the astrometry and image distortion against the Two Micron All Sky Survey (2MASS) catalog.
The image flux levels were done by comparing with the 2MASS catalog.

\subsection{H$_2$ images}
\label{sec:obsh}
The H$_2$ observations were carried out with the WIRCam at the CFHT in July and October 2011.
The H$_2$ narrow-band filter has a bandwidth of 0.032 $\mu$m and a central wavelength of 2.122 $\mu$m.
The seeing was about 0\farcs6 (from 0\farcs5 to 1\farcs0) during these observations.
The targets were observed with five dithered exposures.
An individual exposure of 200\,s was taken in each frame and was repeated 11--16 times for each object.
Again, in order to avoid saturation issues, we took several shorter exposures of 13\,s for the bright objects at NIR wavelength.
The data calibration and reduction are the same as for the Ks-band observations (see Section \ref{sec:obsk}).

\subsection{Archival \textit{Spitzer} data}
We use the archival \textit{Spitzer} InfraRed Array Camera (IRAC) images obtained by the c2d\footnote{http://peggysue.as.utexas.edu/SIRTF/} team \citep{ev03,ev09}.
The IRAC channel 1 and channel 2 (hereafter IRAC1 and IRAC2, respectively) have central wavelength of 3.6 $\mu$m and 4.5 $\mu$m. The spatial resolutions of IRAC1 and IRAC2 are 1\farcs66 and 1\farcs72, and the pixel sizes are 1\farcs22 and 1\farcs21, respectively.
These data were processed through the c2d standard pipeline, as described in the c2d data delivery document \citep{ev07}.

\begin{figure*}
\hspace{-3.1cm}
\includegraphics[scale=.81]{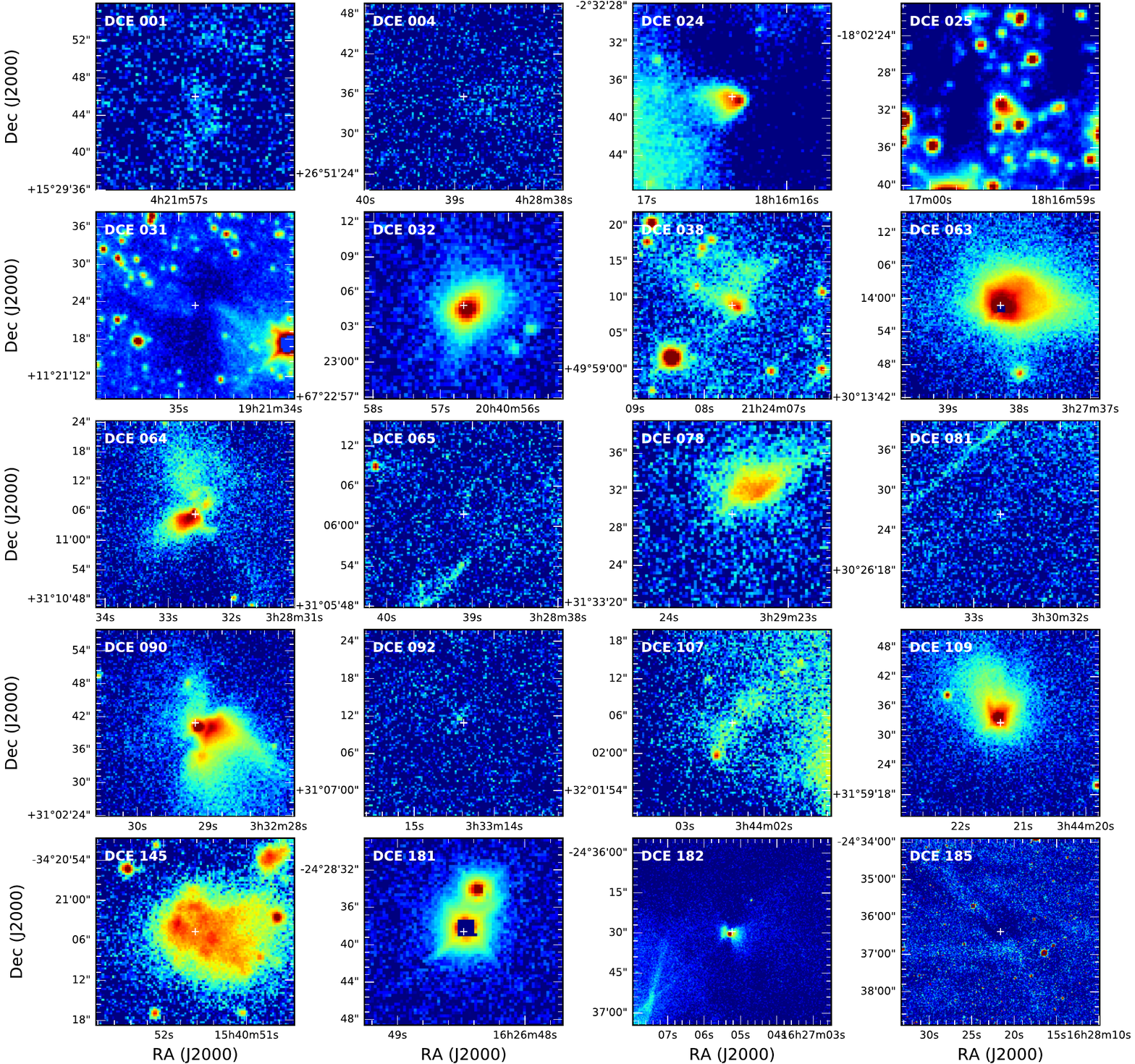}
\caption{Ks-band images toward 20 LLO targets.
The white cross indicates the source position from \citet{du08}.
The images are plotted in logarithmic scale, with different boundaries for each map.
}
\label{fig:kband}
\end{figure*}

\section{RESULTS AND ANALYSIS}
\label{sec:result}
\subsection{Observed images}
\label{sec:obs_r}
\subsubsection{Ks-band and IRAC images}
\label{sec:ksband}
%Ks band detectionsm
Our Ks-band observations reveal a number of sources driving protostellar outflows.
Figure \ref{fig:kband} shows the Ks-band images of the 20 selected LLOs. Thirteen objects are detected (see Table \ref{tab:targets}).
Based on the morphology in the Ks-band image, we identify DCE 145 as a background galaxy misidentified as a VeLLO and exclude it from our sample.
Out of the remaining 12 detected objects, eight sources show a conical or biconical structure (see Figure \ref{fig:kband} and Table \ref{tab:targets}).
The other detected sources show extended emission except for DCE 181.
Although the extended emission may come from a nearly pole-on outflow, we cannot rule out the possibility that it is an infrared nebula or a background galaxy.
In addition, we cannot derive the inclination angle of a nearly pole-on outflow because the inclination angle is estimated based on the contrast between the red- and blue-shifted emissions.
Thus, we exclude those extended objects in our analysis.
As a result, we identify eight outflow-driving LLOs on the basis of the Ks-band data.

%Add IRAC1 data
Because the Ks-band images have a higher spatial resolution than the IRAC1 images, they allow us to explore the outflow structures better.
However, the scattered light at the short wavelength (2.146 $\mu$m) could be greatly attenuated by the circumstellar envelope for deeply embedded protostars.
Since the \textit{Spitzer} IRAC1 images at a longer wavelength are less affected by extinction, they reveal outflow cavities in four additional embedded objects in our sample (DCE 004/L1521F, DCE 001/IRAM 04191, DCE 185/IRAS 16253, and DCE 092).
In addition to extinction, the non-detections in the Ks band can also be explained by nearly edge-on configurations, as indicated by the modeling of the infrared images (see Section \ref{sec:res}).
As a result, outflow cavities are found in 12 LLOs out of 19 on the basis of the Ks-band or IRAC1 images.

%Compare with submillimeter observations
To study the outflows, we compare their morphologies at NIR wavelength with that in CO emission reported in the literature.
In the plane of the sky, the outflow orientations in the NIR observations are approximately consistent with that from the CO observations in DCE 004 (L1521F: Takahashi et al.\ 2012), DCE 025 (L328-IRS: Lee et al.\ 2013), DCE 038 (L1014-IRS: Bourke et al.\ 2005), DCE 185 (IRAS 16253: Stanke et al.\ 2006; Hsieh et al.\ 2016), DCE 092, and DCE 078 (M. Hiramatsu, private communication).
However, the NIR and CO observations could sometimes trace different components of an outflow.
Toward DCE 038, the CO (2--1) map (see Figure 1 in Bourke et al.\ 2005) reveals a compact ($\sim$5\arcsec) bipolar outflow, but our Ks-band image shows a more extended structure ($\sim$10\arcsec) with a large opening angle which is consistent with the H-band and Ks-band images in \citet{hu06}.
Although the CO observations trace large-scale outflows ($\gtrsim$100\arcsec) in DCE 001 (IRAM 04191: Andr\'{e} et al.\ 1999, Lee et al.\ 2002) and DCE 025 (L328-IRS: Lee et al.\ 2013), the infrared images reveal only the inner regions ($\lesssim$10\arcsec) of the outflows.
In addition, the position angle found in the IRAC1 image of DCE 001 differs by $\sim$30 degree from the position angle in the CO observations \citep{an99,le02}.
Besides, the opening angle derived from the CO observations (45$\arcdeg$) is much larger than the angle derived from our fitting (25$\arcdeg$) (see Section \ref{sec:res}).
The reason for this discrepancy remains unclear, and high-angular-resolution, high-sensitivity data at NIR wavelength would be required to understand it.

\begin{figure}
\includegraphics[scale=.46]{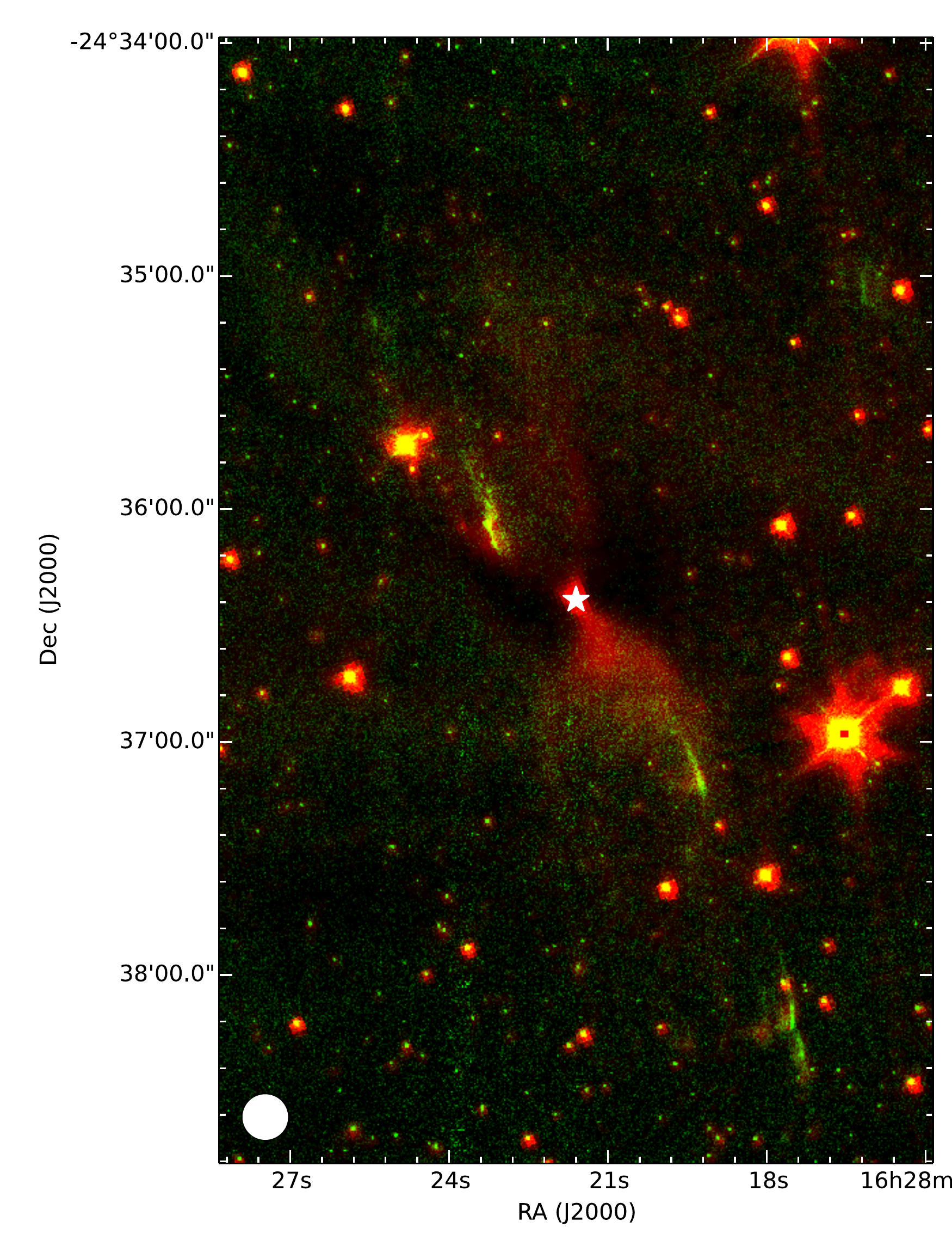}
\caption{Two-color image of DCE 185 with red scale for Spitzer IRAC1 (3.6 $\mu$m) continuum emission and green scale for CFHT H$_2$ emission (2.12 $\mu$m).
}
\label{fig:16253}
\end{figure}

%Ks-band outflow morphologies
Through NIR scattered light, the outflow cavities show quite different morphologies.
The cavity size varies from few hundreds to few thousands au.
Most of the bipolar cavities are asymmetric, being detected (or stronger) in one side only as commonly seen at NIR wavelength due to inclination effects; 
the undetected/weak lobes are believed to be the red-shifted lobes hidden behind the foreground envelopes. 
As a matter of fact, the asymmetries in the infrared images are consistent with the blue/red-shifted lobes detected by CO observations in DCE 001 \citep{an99} and DCE 038 \citep{bo05}.
CO blue-shifted and red-shifted emissions appear in both sides of the outflow in DCE 004 \citep{ta13} and DCE 025 \citep{le13}, suggesting a nearly edge-on inclination and a probably large opening angle; this result is compatible with the NIR observations.
However, for DCE 185, the brighter NIR lobe in the south-west overlaps with the red-shifted lobe found by multitransition CO observations \citep{hs16}.
We speculate that this inconsistency is due to an intrinsic property;
instead of inclination effects, the south-west lobe may contain more scattered light because of heating by the UV-radiation from the stronger south-west H$_2$ jet (Figure \ref{fig:16253}).

This discussion reveals that the NIR asymmetric bipolar cavities provide clues to estimate the inclination angle of the outflows.
In most cases, the bipolar cavities would appear more symmetric in the systems with smaller inclination angles (with respect to the plane of the sky) and would be more asymmetric in those with larger inclination angles.
Thus, we use this property to derive the inclination angle by modeling the observed images (see Section \ref{sec:model})

\begin{figure*}
\hspace{-3.1cm}
\includegraphics[scale=.81,angle=90]{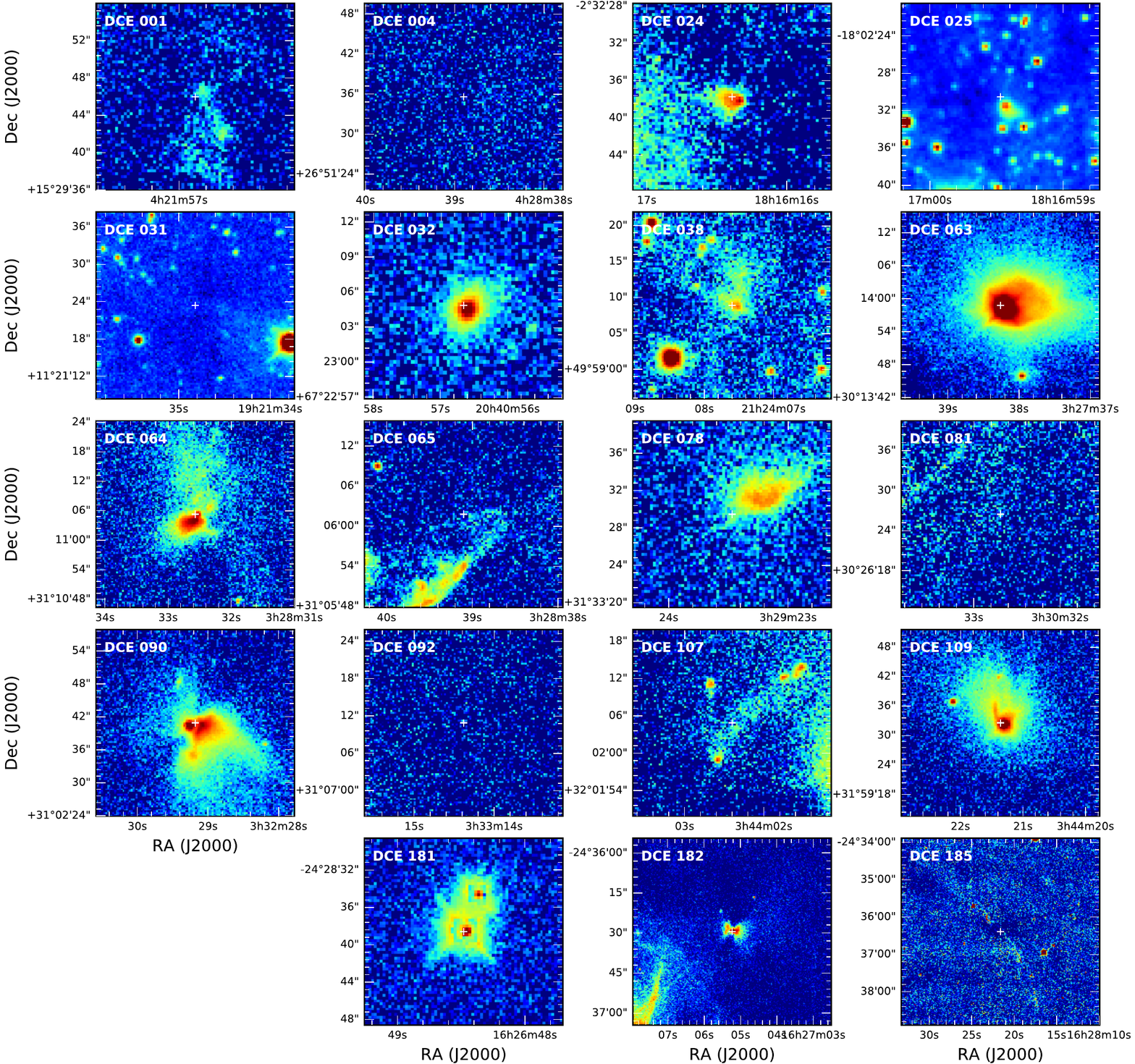}
\caption{Same as Figure \ref{fig:kband} but for H$_2$ images toward 19 LLO targets.
The images are plotted in logarithmic scale, with different boundaries for each map.
}
\label{fig:hband}
\end{figure*}

\subsubsection{H$_2$ images}
Figure \ref{fig:hband} shows the H$_2$ images toward the 19 LLOs (DCE 145 has been excluded, see Section \ref{sec:ksband}).
The H$_2$ observations were performed with a narrow filter but with a long integration time, allowing them to trace the continuum emission as well as the Ks-band image.
The H$_2$ images resembling the Ks-band ones are likely dominated by the continuum emission (Figure \ref{fig:hband}).
Hereafter, we call Figure \ref{fig:hband} as ``H$_2$ image(s)'' when it actually consist of H$_2$ emission and continuum emission.  
We note that the Ks broad-band filter includes the H$_2$ line, but with a 10 times larger filter width corresponding to $\sim$45000 km s$^{-1}$. 
Thus, the H$_2$ emission is greatly diluted in the Ks-band image.

Comparing with the Ks-band images, our H$_2$ survey reveals a robust jet only toward DCE 185 (IRAS 16253, Figure \ref{fig:16253}) as in the 2.12 $\mu$m H$_2$ image of \citet{kh04} and in the mid-infrared InfraRed Spectrometer (IRS) H$_2$ image of \citet{ba10}.
The H$_2$ jet of DCE 185 shows a prominent ``S-shaped'' symmetry around the central source (Figure \ref{fig:16253}). 
It is believed to originate from the tidal interactions between the disk where the jet originates and a companion in a noncoplanar orbit \citep{hs16}.

Besides, we identify three sources with marginal H$_2$-jet detections by eye.
Because the H$_2$ image can be dominated by the continuum emission, we search for jet-like structures that are seen in the H$_2$ image (Figure \ref{fig:hband}) but not in the Ks-band image (Figure \ref{fig:kband}).
As a result, we find three H$_2$-jet driving candidates: DCE 001, 109, and 107.
These sources show very different morphologies of jets:
(1) DCE 001 shows very weak H$_2$ emission roughly along the outflow direction seen in the CO (2--1) map from the IRAM 30 m telescope \citep{an99}. 
(2) In DCE 109, a collimated H$_2$ emission appears from the central star to the north and may extend to a clear H$_2$ jet knot. 
(3) We found two H$_2$ knots in the north-west direction that are likely driven by DCE 107 along a collimated jet.
Because DCE 001, 109, and 107 have only marginal detections of H$_2$ jets, we call these three objects ``jet-driving candidates''.

%Statistics % detected frequency, detected =>?, undetected =>?
Statistically, our H$_2$ jet survey has a low detection rate: only one prominent H$_2$ jet-driving source and three jet-driving candidates out of 19 LLOs. 
These three candidates, if real at all, show very weak H$_2$ emission.
Because the H$_2$ line emission usually traces high-velocity gas ($>$100 km s$^{-1}$) that produces high-energy photons exciting H$_2$ \citep{wo91},
this result implies that most of the LLOs drive low-velocity outflows rather than high-velocity jets.
Therefore, the low detection rate of H$_2$ jets is consistent with a scenario in which LLOs are at a quiescent accretion phase and thus drive weak outflows.

\subsection{Infrared image modeling}
\label{sec:model}
Measuring the outflow opening angle requires a good knowledge of the inclination angle because, due to inclination effects, the apparent opening angle seen in the image is in fact larger than the true opening angle. 
Inclination angles can be estimated based on the intensity ratio between the blue-shifted and red-shifted lobes. 
Considering a spherically-symmetric envelope, the scattered light from the red-shifted lobe is more attenuated, by the thick foreground envelope, than that from the blue-shifted lobe.
Therefore, a large inclination angle yields a large difference in brightness between the blue-shifted and red-shifted outflows, which allows us to derive the inclination angle, and in turn the opening angle.

We fit the NIR images (Ks-band or IRAC1) with the radiative transfer model of \citet{wh03a} for the 12 sources with outflow detections (see Section \ref{sec:ksband} and Table \ref{tab:targets}).
Note that although we tried to fit the three extended sources (Table \ref{tab:targets}), the best-fit results were very different from the observations (see Section \ref{sec:ksband}). Thus, we exclude these three sources in our analysis.
We remind the reader that the IRAC1 images have a higher sensitivity but a much lower angular resolution (see Section \ref{sec:ksband}).
Thus, the IRAC1 images enable us to probe outflows from more embedded objects, and the Ks-band images provide better constraints on the opening angles.

\subsubsection{Setup of the models}
\label{sec:setup}
%Whitney's code
\citet{wh03a} provide a radiative transfer code to model the scattered light images of an outflow cavity (hereafter Whitney's code/model).
This code is based on the Monte Carlo radiative equilibrium routine developed by \citet{bj01}, which calculates the radiative transfer in a three-dimensional spherical-polar grid.
Whitney's code models a spherically-symmetric envelope with biconical cavities carved out by a bipolar outflow viewed at different inclinations.
The default cavity shape follows $z\propto\sqrt{x^2+y^2}^b$ where $x$, $y$, and $z$ are the Cartesian coordinates and $b$ is the power-law exponent of 1.5.
Using this code, \citet{wh03b} present an evolutionary sequence of models for a low mass protostar at stages of Class 0, Late 0, I, Late I, II, and III.
Because our targets have bolometric temperatures ranging from 20 K to 114 K (Table \ref{tab:targets}),
we take the Class Late 0 model as the template model and then vary the opening angle and inclination angle to fit the observed Ks-band and IRAC1 images.

% Parameter setting
% Several tests 1. EXF for rho^exf in cavity, 2. Lum of star, 3. rho constant, 4. Minfall
Here, we discuss how the parameter set in Whitney's code may affect the modeled images.
Whitney's code categorizes several tens of parameters into three groups: central star, disk, and envelope.
We speculate that the central star and disk may only weakly affect the large-scale outflow cavities at hundreds to thousands au, when
the disk of a Class Late 0 source \citep{wh03b} has an outer radius of 50\,au ($\sim$0\farcs2--0\farcs4 in our sample).
In addition, to remove the influence of the star and disk emission, we mask the central source in a circular region in our fitting process (see Section \ref{sec:res}). 
Furthermore, we find that the internal luminosity (associated to parameters of star and disk) likely affects only the image brightness scale of the cavity but not the distribution or structure.
Therefore, we use a linear function to scale the intensity map (see Section \ref{sec:res}), which makes our fitting results independent of the internal luminosity. 
On the other hand, we test whether or not the envelope properties could change the brightness distributions of the outflow cavities.
We find that the density profiles in the cavity can significantly change the intensity scale but not the distribution.
We also find that the mass infall rate ($\dot{M}_{\rm env}$) can alter the intensity distribution but does not significantly affect the opening angle, and for the Class Late 0 objects, it is set to 10$^{-5}$ $M_\odot\rm~yr^{-1}$ \citep{wh03b}.

\begin{figure*}
\includegraphics[scale=.54]{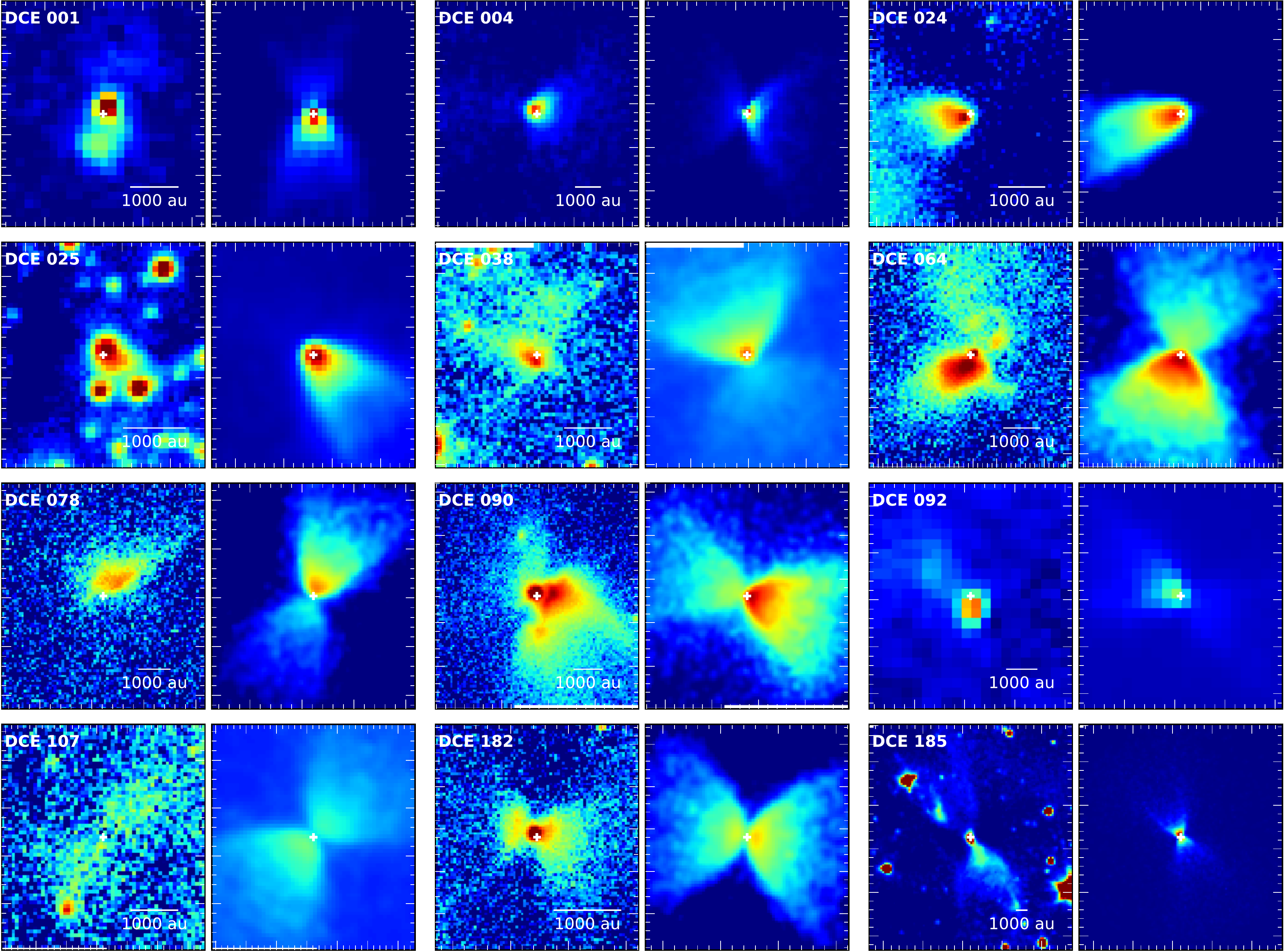}
\caption{Comparison of observed images (left panel of each pair) with best-fit modeled images (right panel of each pair).
The images of DCE 004, 001, 185, and 092 are the IRAC1 images and the rest are the Ks-band ones.
}
\label{fig:mod}
\end{figure*}

\subsubsection{Fitting the opening angles and inclination angles}
\label{sec:res}
% modeling
%We derive the outflow opening angles and inclination angles by fitting the observed images with the modeled images from \citet{wh03a}. 
%An accurate measurement of the opening angle relies on a good constraint of the inclination angle because the true opening angle is in fact smaller than that seen in the image due to the inclination effect.

There are several steps in our fitting process.
Since the luminosity of the central source can change the image brightness, we linearly scale the intensity of the modeled image to fit the observations.
To remove the effects of field stars and the central object (star$+$disk), we mask these sources with adequate circles determined by eye.
Using Whitney's code, we construct a grid of models that vary opening angle ($\theta_{\rm open}$), inclination angle ($\theta_{\rm inc}$), and position angle (P.A.) with a cell size of 5$\arcdeg$ for all dimensions.
Then, we derive the $\chi^2$ values between the observed image and each model, and as a result, we obtain a $\chi^2$ distribution in the three-dimensional grid.
Then, we find the best-fit opening angle and inclination angle, and taking a 99.7\% confidence level with five free parameters ($\theta_{\rm open}$, $\theta_{\rm inc}$, P.A., and two in the linear function) ($\Delta\chi^2>18.2$, Press et al.\ 1992), we obtain its upper and lower limits (Table \ref{tab:targets}).
Figure \ref{fig:mod} shows the images of the best-fit models and the corresponding observed images.

\begin{figure*}
%\vspace{100pt}
\includegraphics[scale=.6]{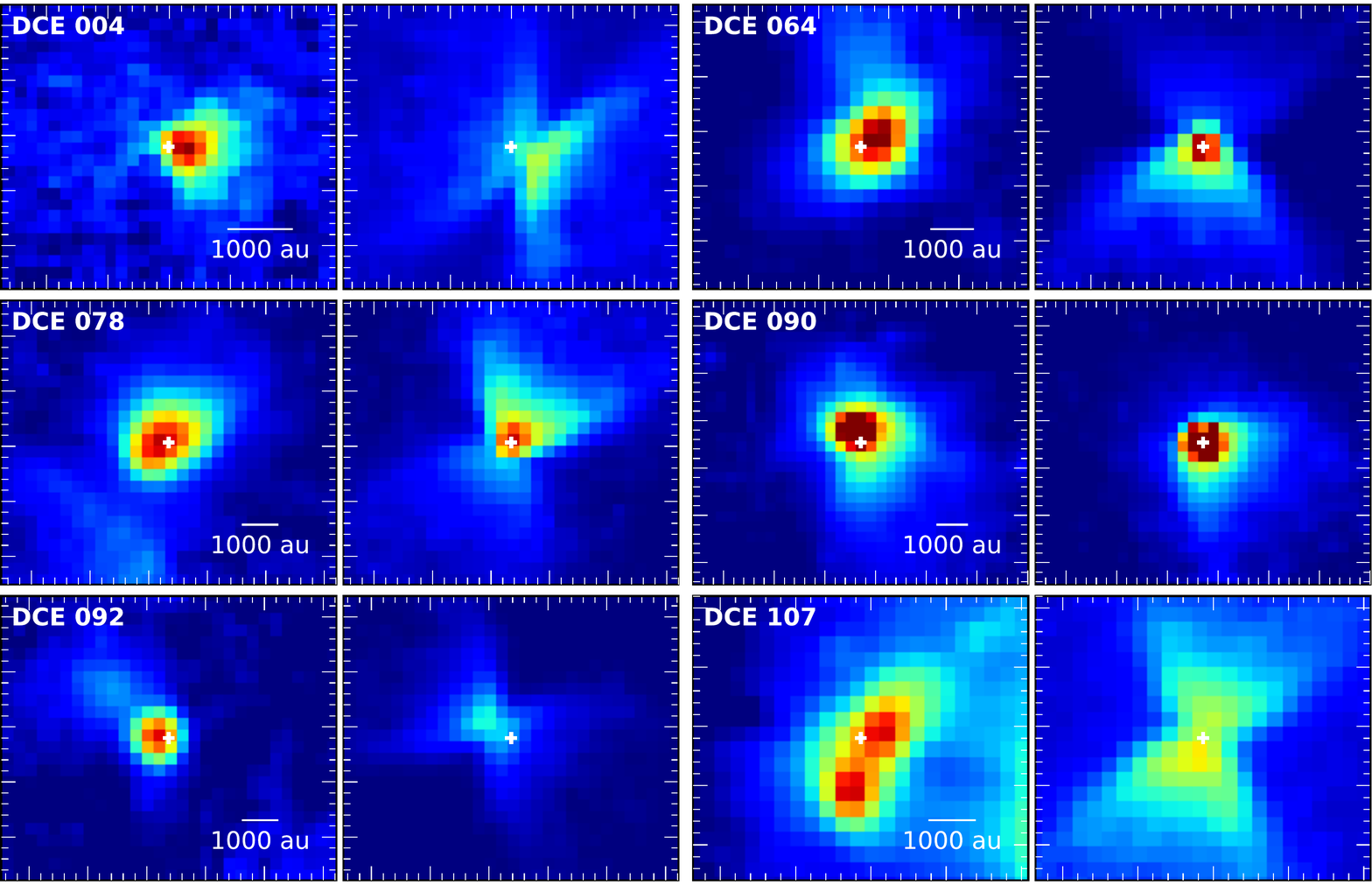}
\vspace{-20pt}
\caption{ Same as Figure \ref{fig:mod} but with the IRAC1 images (DCE 064, 078, 090, and 107) or the IRAC2 images (DCE 004 and 092).
Each pair shows the observed IRAC1 or IRAC2 image (left panel) and the best-fit modeled image (right panel).
}
\label{fig:mod4}
\end{figure*}

We discuss here two special cases, DCE 185 and 092, with arbitrary decisions in our fitting process.
For DCE 185, as the south-west lobe is brighter than the north-east lobe in the IRAC1 image (Figure \ref{fig:16253}),
this brightness asymmetry is in conflict with the blue- and red-shifted outflows identified by CO observations \citep{st06,ma13,hs16}.
Thus, we speculate that this asymmetric bipolar cavity is due to an intrinsic property rather than inclination effects (see Section \ref{sec:ksband}).
In the fitting, we restrict the red-shifted lobe to be in the south-west as seen in the CO observations, and since the model assumes the asymmetry comes from inclination effects,
we obtain an inclination angle of 5\arcdeg, the lower limit of our model grid.
This underestimate of the inclination angle yields an overestimate of the opening angle.
Therefore, we take the derived opening angle of DCE 185 as an upper limit for later analysis.
Besides, the $\chi^2$ distribution of DCE 185 shows two local minima at opening angles of 125$\arcdeg$ and 35$\arcdeg$.
We eliminate the larger opening angle, since it is most likely caused by contamination of diffuse cloud emission.
For DCE 092, the $\chi^2$ distribution also has two local minima located at opening angles of 100$\arcdeg$ and 35$\arcdeg$.
We eliminate the larger opening angle because it is likely affected by the nearby bright sources.

\begin{deluxetable}{ccccccc}
\tablewidth{0pt}
\tabcolsep=0.06cm
\tabletypesize{\tiny}
\tablecaption{Results of fits to Ks-band and IRAC1 data.}
\tablehead{ 
\colhead{}    & \multicolumn{2}{c}{Ks-band} & \multicolumn{2}{c}{IRAC1} & \multicolumn{2}{c}{ IRAC2} \\
\cmidrule(lr){2-3} \cmidrule(lr){4-5} \cmidrule(lr){6-7}
\colhead{Source}		& \colhead{Op. angle} & \colhead{Inc. angle}	& \colhead{Op. angle} & \colhead{Inc. angle} & \colhead{Op. angle} & \colhead{Inc. angle}\\
\colhead{} & \colhead{(deg}		& \colhead{(deg)}	& \colhead{(deg)}		& \colhead{(deg)}	& \colhead{(deg)}		& \colhead{(deg)}
}
\startdata 
DCE 004		& -	& -	& 60$^{+2.5}_{-7.5}$	& 15$^{+2.5}_{-2.5}$	& 70$^{+2.5}_{-2.5}$	& 10$^{+2.5}_{-2.5}$	\\
DCE 064			& 55$^{+2.5}_{-2.5}$	& 20$^{+2.5}_{-2.5}$	&	45$^{+2.5}_{-2.5}$	& 20$^{+2.5}_{-2.5}$		& -	& -\\
DCE 078			& 40$^{+2.5}_{-2.5}$	& 20$^{+2.5}_{-2.5}$	& 40$^{+2.5}_{-2.5}$	& 20$^{+2.5}_{-2.5}$	& -	& -\\
DCE 090			& 45$^{+2.5}_{-2.5}$	& 25$^{+2.5}_{-2.5}$ &	30$^{+2.5}_{-2.5}$	& 55$^{+2.5}_{-2.5}$		& -	& -\\      
DCE 092	 	& -	& -	& 25$^{+12.5}_{-2.5}$	& 15$^{+7.5}_{-12.5}$	& 45$^{+2.5}_{-17.5}$	& 5$^{+7.5}_{-2.5}$ \\
DCE 107			& 55$^{+2.5}_{-27.5}$& 5$^{+2.5}_{-2.5}$		& 40$^{+2.5}_{-2.5}$	& 5$^{+2.5}_{-2.5}$	& -	& -
\enddata
\label{tab:irac}
\end{deluxetable}

\begin{figure}
\includegraphics[scale=.43]{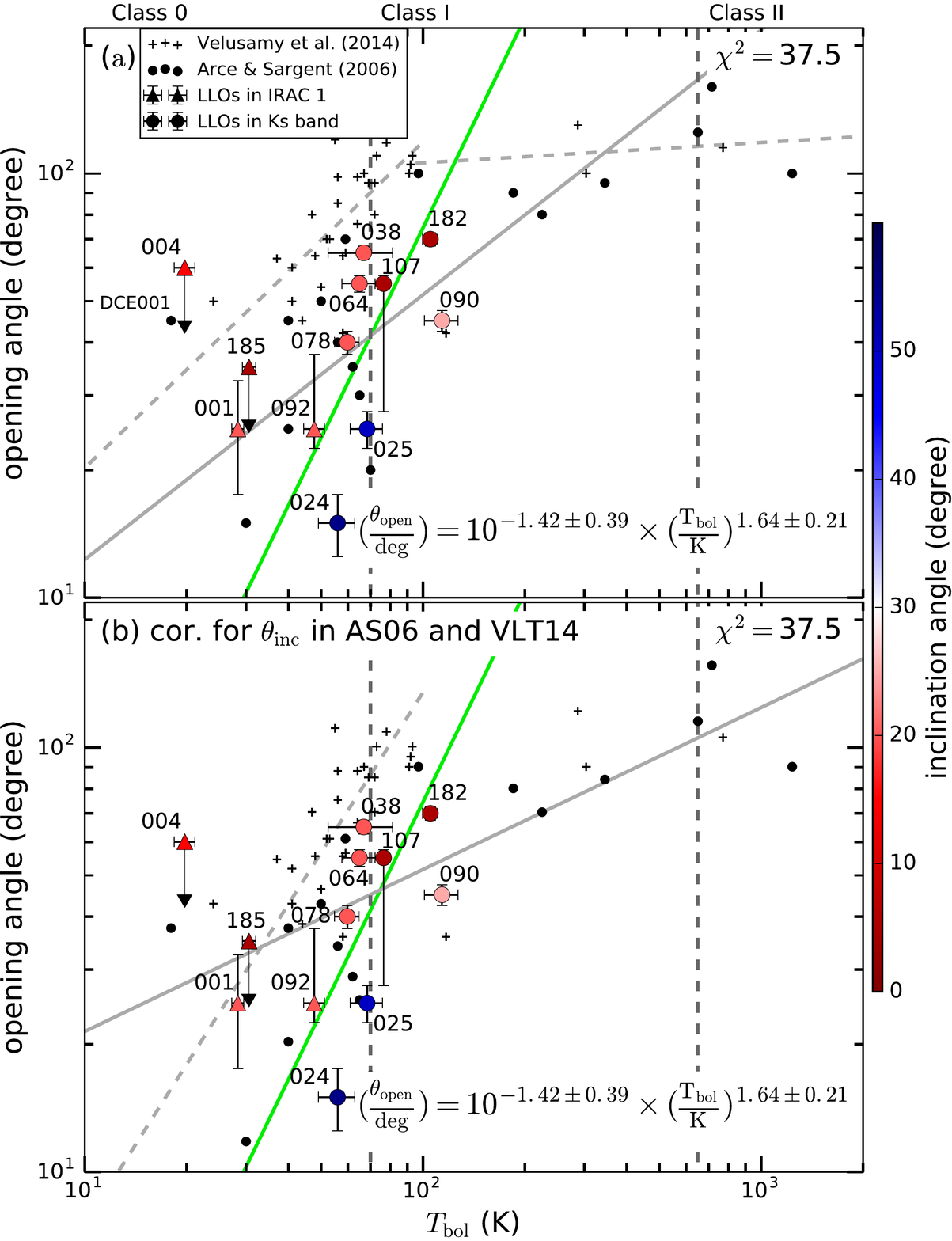}
\vspace{-20pt}
\caption{
(a) Plots of outflow opening angle versus bolometric temperature. (b) Same as (a) but with inclination correction for opening angles in AS06's and VLT14's samples (assuming $\theta_{\rm inc}=32\arcdeg7$).
The color circles (triangles) indicate the LLOs with the opening angle obtained from Ks-band (IRAC1 band).
The vertical dashed lines show the boundaries of Class 0, I and II in bolometric temperatures.
The black dots and plus signs show the sources of AS06 and VLT14, and the respective gray solid and gray dashed lines indicate the correlations derived by these two studies (Equations \ref{eq:arce2} and \ref{eq:vela}).
The green line shows our best-fit power-law for LLOs. The parameters of the best fit are written in the bottom right corner} (Equation \ref{eq:llo}).
\label{fig:wid}
\end{figure}

To check our fitting results,
we fit the IRAC1 images in four sources (DCE 064, 107, 090, and 078) that have outflow detections at IRAC1 in addition to the Ks-band,
and the IRAC2 images of the following two sources: (1) DCE 004, which is an outlier in the $T_{\rm bol}-\theta_{\rm open}$ plot (see Section \ref{sec:opeTbol} and Figure \ref{fig:wid}), and (2) DCE 092, for which the inclination angle derived from the fitting of its IRAC1 image has a large uncertainty.
Comparing the fitting results at two wavelengths, we obtain consistent inclination angles and similar opening angles (difference $\leqslant$ 10\arcdeg~or within the errors), except for DCE 090 (Table \ref{tab:irac} and Figure \ref{fig:mod4}).
In DCE 090, we suspect that the discrepancy between $\theta_{\rm inc, IRAC1}=55\arcdeg$ and $\theta_{\rm inc, Ks}=25\arcdeg$ comes from an unreliable fitting in IRAC1 due to the contamination from the bright central object.
Because the Ks-band image has a better resolution, we believe that the result from the Ks-band is more reliable for DCE 090.
Overall, we conclude that fitting the images at different wavelengths gives comparable results.
As a result, we list the best-fit results from the relatively shorter wavelengths in Table \ref{tab:targets} and use these values in the following analysis.

\subsubsection{Fitting with streamline cavity}
%For precessing jet see Whitney 2003a
In addition to the curved cavity mentioned above, Whitney's code provides an alternative cavity shape namely streamline which is conical on large scales and might be carved out by precessing jets \citep{wh03a}.
To study the origin of the outflow widening, we also fit the observed images with the model of streamline cavity and compare the results with those obtained assuming a curved cavity.
We find that, except for DCE 001 and DCE 024 that are better fitted with the streamline cavity, all other objects are better fitted with the curved cavity above a confidence level of 99.7\%.
DCE 001 and DCE 024 have the smallest opening angles and relatively low bolometric temperatures among our targets, suggesting that they are younger than other sources. 
Although, based on this, one could be tempted to conclude that the outflow cavity is 
carved by precessing jets at the earliest stage, a sample of only two sources is not large enough to be statistically meaningful.
In addition, the large-scale CO outflows in DCE 001 do not show the streamline shape \citep{an99}.
Therefore, we lack evidence to support the fact that the outflow widening is caused by carving of precessing jets.

\section{DISCUSSION}
\label{sec:discussion}
\subsection{Outflow opening angle versus bolometric temperature}
\label{sec:opeTbol}
%Consistent?
% separate into VeLLOs and LLOs? or separate LLOs with an opening angle of 30 degree?
%\subsection{\color{blue} opening angle compare with other par}
Here, we compare our LLOs with YSOs from AS06 and VLT14 in the plot of bolometric temperature versus opening angle (Figure \ref{fig:wid}a).
To study the evolution of protostellar outflows, previous works have studied the correlation between outflow opening angle and bolometric temperature taken as an evolutionary indicator.
Based on their survey of CO outflows toward YSOs,
AS06 found that the outflow opening angle widens as the core evolves from Class 0 to Class III.
They found a correlation between the outflow opening angle ($\theta_{\rm open}$) and the bolometric temperature ($T_{\rm bol}$) as
%\begin{equation}
%\log(\frac{\theta_{\rm open}}{\rm deg}) = (0.7 \pm 0.2)+(0.26 \pm 0.04)\log(\frac{t}{\rm yr})
%\label{eq:arce}
%\end{equation} 
%where $t$ is obtained using the $T_{\rm bol}$-age relation 
%\begin{equation}
%$\log(\frac{t}{\rm yr})=2.4\times\log(T_{\rm bol})-0.9$
%\label{eq:lada}
%\end{equation} 
%from \citet{la98}, and thus the original $T_{\rm bol}$-$\theta_{\rm open}$ correlations is
\begin{equation}
\log(\frac{\theta_{\rm open}}{\rm deg}) = (0.47 \pm 0.20)+(0.60 \pm 0.10)\log(\frac{T_{\rm bol}}{\rm yr}).
\label{eq:arce2}
\end{equation}
Lately, using {\it Spitzer} IRAC images, VLT14 measured the outflow opening angles toward 31 YSOs and obtained a $T_{\rm bol}-\theta_{\rm open}$ correlation of
\begin{subequations}
\label{eq:vela}
\begin{align}
\log(\frac{\theta_{\rm open}}{\rm deg}) & = 0.54+0.77\log(\frac{T_{\rm bol}}{\rm yr}),~{\rm for}~T_{\rm bol}<100K\\
							&= 1.92+0.05\log(\frac{T_{\rm bol}}{\rm yr}),~{\rm for}~T_{\rm bol}>100K. 
\end{align}
\end{subequations}
Before we compare our sample with AS06's and VLT14's results, two caveats should be made.
First, because both AS06 and VLT14 measured the opening angles without inclination correction, their measurements should be viewed as upper limits in comparison with ours.
Second, the NIR scattered light could highlight a broader outflow cavity than CO \citep{oh97,ta97,le06,la08,ar13}, because CO may not be able to trace the full extent of the NIR reflection nebula due to its high opacity especially at the velocities close to the ambient cloud velocity \citep{ar13}.

We find that most LLOs fit well the $T_{\rm bol}-\theta_{\rm open}$ trend found by AS06, except for DCE 004 (L1521F) and DCE 024 (CB-130-1-IRS1).
Two reasons may explain the deviation of DCE 004:
(1) DCE 004 may host a binary system driving two outflows, as SMA CO (2--1) observations suggest \citep{ta13}. The possible two outflows have different axes as seen in many other cases \citep{of16,le16}, which can affect our opening angle estimate with NIR observation.
Therefore, we consider the measured opening angle as an upper limit.
(2) Our fitting result implies a nearly edge-on configuration, which may cause the low bolometric temperature (see Section \ref{sec:corT}).
Although DCE 024 deviates from the power-law relation of AS06, it fits well the new power-law index that better describes the sources with $T_{\rm bol}\lesssim 120$ K (see below).

%In order to compare with Arce \& Sargent's results, we
We use a single power-law to fit our LLO sample (Figure \ref{fig:wid}a). The best fit is
\begin{equation}
(\frac{\theta_{\rm open}}{\rm deg}) = 10^{-1.42\pm0.39} \times (\frac{T_{\rm bol}}{K})^{1.64\pm0.21}.
\label{eq:llo}
\end{equation}
We have excluded DCE 004 and DCE 185 from the fitting since their fitted opening angles are considered as upper limits.
The best-fit power-law index, 1.64, is much larger than the index derived by AS06.
If we include the data points with $T_{\rm bol} \leqslant120$ K from AS06, we obtain $(\frac{\theta_{\rm open}}{\rm deg}) = 10^{-1.02\pm0.26} \times (\frac{T_{\rm bol}}{K})^{1.46\pm0.14}$, which is consistent with Equation \ref{eq:llo} within $1\sigma$.
This suggests that we need a broken power law or a more complicated model to describe how the outflow cavities widen.
In addition, based on this large index, we suggest that the outflow opening angle may better discriminate the evolutionary state of protostars than the bolometric temperature in a range of $T_{\rm bol} \leqslant 120~K$.

VLT14 used two power-law components to fit their data and found a break at $T_{\rm bol} \approxeq 100~K$. However, the steeper power law at the lower $T_{\rm bol}$ end (Equation \ref{eq:vela}a) is quite different from our fitting result (Equation \ref{eq:llo}). We speculate that this discrepancy partially comes from inclination effects that were not taken into account by VLT14. 
To correct for this, we assume that all sources of VLT14 (as well as AS06) have an inclination angle of 32\arcdeg7 (with respect to the plane of the sky) and we estimate their inclination-corrected opening angles (Figure \ref{fig:wid}b); the angle of 32\arcdeg7 corresponds to a mean inclination angle assuming all orientations are equally favorable \citep[note that the angle 57\arcdeg3 in the references is with respect to the line of sight]{bo96,du14}.
After the correction, the VLT14 sample is broadly consistent with our LLOs.
We find a new-best fit of $\log(\frac{\theta_{\rm open}}{\rm deg}) =-0.38+1.25\log(\frac{T_{\rm bol}}{\rm yr})$ for $T_{\rm bol}<100K$ for the inclination-corrected sample of VLT14, which is consistent with our LLOs (Equation \ref{eq:llo}) within $\sim$3-4$\sigma$.

We now compare the opening angles of our LLO sample with the Class 0 samples studied by AS06, \citet{se08}, and VLT14.
The AS06 sample includes 11 Class 0 objects, Seale \& Looney's sample has 21 Class 0 objects, and the VLT14 sample contains 20 Class 0 objects.
Out of the 12 LLOs with NIR outflow detections in our sample, five ($\sim$40\%) have opening angles smaller than $25\arcdeg$. Such small opening 
angles are rare in the Class 0 samples of AS06, \citet{se08}, and VLT14 (3/11, 2/21, and 0/20, respectively).
Note that the ratios remain the same after the inclination correction.
Assuming that the opening angle is an evolutionary indicator, we conclude that our sample of outflow-driving LLOs contains a higher 
fraction of young objects than the samples of AS06, \citet{se08}, and VLT14.
In turn, this suggests that at least the five LLOs with $\theta_{\rm open}\leqslant25\arcdeg$ may be very young Class 0 objects.

\subsection{Approximate correction for inclination effects on $T_{\rm bol}$}
\label{sec:corT}
Figure \ref{fig:wid}a shows that the LLOs below the best-fit line have relatively large inclination angles while the sources above have relatively small inclination angles. Because the inclination angle can affect the Spectral Energy Distribution (SED), and in turn the bolometric temperature \citep{wh03a,wh03b,ro06,cr08}, 
this geometrical effect may cause the LLOs to deviate from the best-fit relation.

\begin{figure}
\includegraphics[scale=.43]{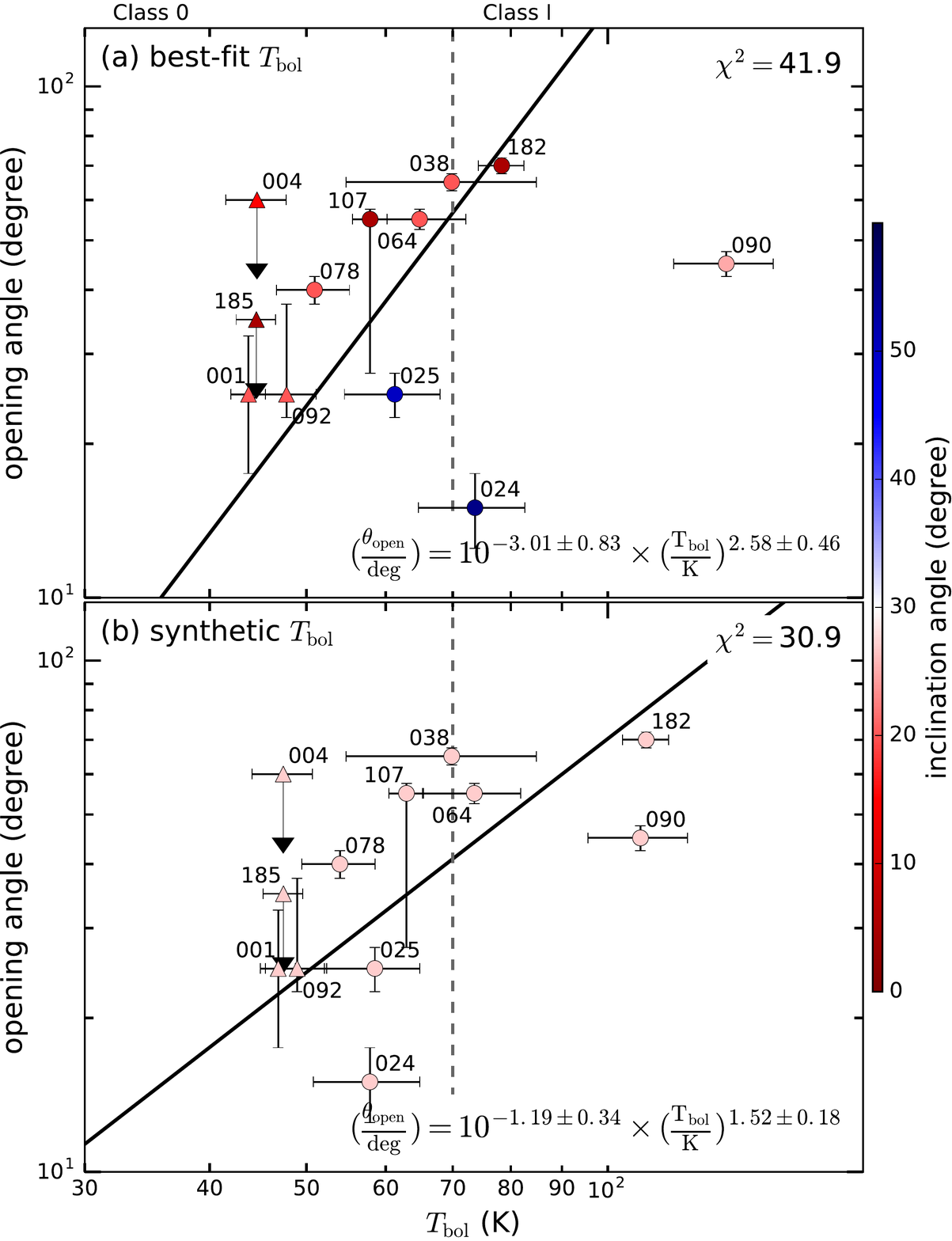}
\vspace{-20pt}
\caption{
(a) Same as Figure \ref{fig:wid} but using $T_{\rm bol}$ from the best-fit modeled SED in LLOs. (b) Same as (a) but using $T_{\rm bol}$ from the synthetic SED with $\theta_{\rm inc}=27\arcdeg$ (see Section \ref{sec:corT}).
The black line shows our best-fit power-law.}
\label{fig:wid2}
\end{figure}

To investigate whether the inclination effects cause the dispersion in Figure \ref{fig:wid}a, we correct the bolometric temperatures for inclination effects using the SED fitting tool of \citet{ro06,ro07}.
This SED fitting tool, constructed with the radiative transfer code of \citet{wh03a}, consists of 20,000 YSO models with SEDs computed at 10 viewing angles for each model, resulting in 200,000 SEDs in total.
%(The ten viewing angles are 3$\arcdeg$, 9$\arcdeg$, 14$\arcdeg$, 20$\arcdeg$, 27$\arcdeg$,  33$\arcdeg$,  41$\arcdeg$,  49$\arcdeg$,  58$\arcdeg$,  72$\arcdeg$).
Since the SEDs' uncertainty grows rapidly at wavelengths $\gtrsim$100 $\mu$m and becomes very high at wavelengths $\gtrsim$600 $\mu$m \citep[see Figure 1 in][]{ro06}, we ignore the SED data points at wavelengths $>$850 $\mu$m, and for the sources lacking 850 $\mu$m data, interpolate a flux at 850 $\mu$m logarithmically.
From the processed SEDs, we calculate the bolometric temperatures (Table \ref{tab:targets}). The derived temperatures are consistent with the results of \citet{du08} within the uncertainties except for DCE 031 (DCE 031 has no outflow detection such that it will not affect the following analysis.). 
We fit the processed SEDs with the YSO models of \citet{ro06,ro07}.
Then, we identify the best-fit model as the one with the lowest $\chi^2$ and with the consistent $\theta_{\rm open}$ and $\theta_{\rm inc}$ from our image fits (see Section \ref{sec:res}) within a difference $<$10\arcdeg. 
To get rid of inclination effects on $T_{\rm bol}$, we define an inclination-independent bolometric temperature as the bolometric temperature of the synthetic SED of the best-fit model computed for an arbitrary inclination angle of 27\arcdeg. This inclination angle corresponds to about the averaged value in our sample (Table \ref{tab:targets}).
We remind the readers that our survey at NIR wavelength cannot identify outflows at a nearly pole-on configuration. Therefore, a small inclination angle is reasonable.
We calculate the bolometric temperatures from the best-fit SED and from the synthetic SED, namely best-fit $T_{\rm bol}$ and synthetic $T_{\rm bol}$, respectively.
With the same inclination, we can use the synthetic $T_{\rm bol}$ as an evolutionary indicator without inclination effects.
Figures \ref{fig:wid2}a and \ref{fig:wid2}b show the $T_{\rm bol}-\theta_{\rm open}$ relations with the best-fit $T_{\rm bol}$ and synthetic $T_{\rm bol}$, respectively, in which, for comparison, we assign the same uncertainty as the observed $T_{\rm bol}$ in percentage.
By fitting a single power-law as Equation (\ref{eq:llo}), we find that the $T_{\rm bol}-\theta_{\rm open}$ correlation with the synthetic $T_{\rm bol}$ has a smaller $\chi^2$ (see Figure \ref{fig:wid2}).
This result supports our speculation that the inclination effects may cause the deviation from the best-fit $T_{\rm bol}-\theta_{\rm open}$ relation.

There are three caveats about the inclination correction.
First, although we use the image fits to constrain $\theta_{\rm open}$ and $\theta_{\rm inc}$, the SED fit may still be degenerated, preventing us from determining the physical parameters.
Thus, our inclination correction on $T_{\rm bol}$ should be considered as a rough approximation rather than a precise determination.
Second, the SED models of \citet{ro06} do not include brown dwarfs while they are used to interpret the low luminosity of VeLLOs in the literature \citep{bo05,ka11,le13,pa14}.
Third, the synthetic $T_{\rm bol}-\theta_{\rm open}$ correlation is probably a self-consistent result produced artificially by the radiative transfer code \citep{wh03a, ro06, ro07} because both the synthetic $T_{\rm bol}$ and $\theta_{\rm open}$ correspond to the model rather than the observation.
%{\color{blue} To test whether the synthetic $T_{\rm bol}$ can better indicate the evolutionary stage, we study the $T_{\rm bol}-L_{\rm bol}/L_{\rm smm}$ correlations using the best-fit $T_{\rm bol}$ and synthetic $T_{\rm bol}$ (Figure \ref{fig:TbolLL}).}
Despite these caveats, the synthetic $T_{\rm bol}-\theta_{\rm open}$ and the observed $T_{\rm bol}-\theta_{\rm open}$ have very similar power-law fits (see equations in Figures \ref{fig:wid} and \ref{fig:wid2}b).
This implies that the radiative transfer code reproduces well the observed $T_{\rm bol}-\theta_{\rm open}$ correlation at least for these LLOs.
This discussion tentatively suggests that the tighter synthetic $T_{\rm bol}-\theta_{\rm open}$ correlation is reliable.

To further test the inclination correction on $T_{\rm bol}$, we compare the best-fit $T_{\rm bol}$ and the synthetic $T_{\rm bol}$ with the ratio of bolometric to submillimeter luminosity ($L_{\rm bol}/L_{\rm smm}$)  taken as a better evolutionary indicator. We exclude DCE 038 because of its problematic measurement of $L_{\rm bol}/L_{\rm smm}$ (see Section \ref{sec:LL}).
We find that $L_{\rm bol}/L_{\rm smm}$ is slightly better correlated to the synthetic $T_{\rm bol}$ with $r=0.57$ and $\rho=0.69$ (p-values of 0.11 and 0.04, respectively) than to the best-fit $T_{\rm bol}$ with $r=0.09$ and $\rho=0.50$ (p-values of 0.81 and 0.17, respectively).
Although the sample is small, this result suggests that the synthetic $T_{\rm bol}$ could better reflect the evolutionary stage than the best-fit $T_{\rm bol}$.

\begin{figure}
\vspace{-70pt}
\includegraphics[scale=.41]{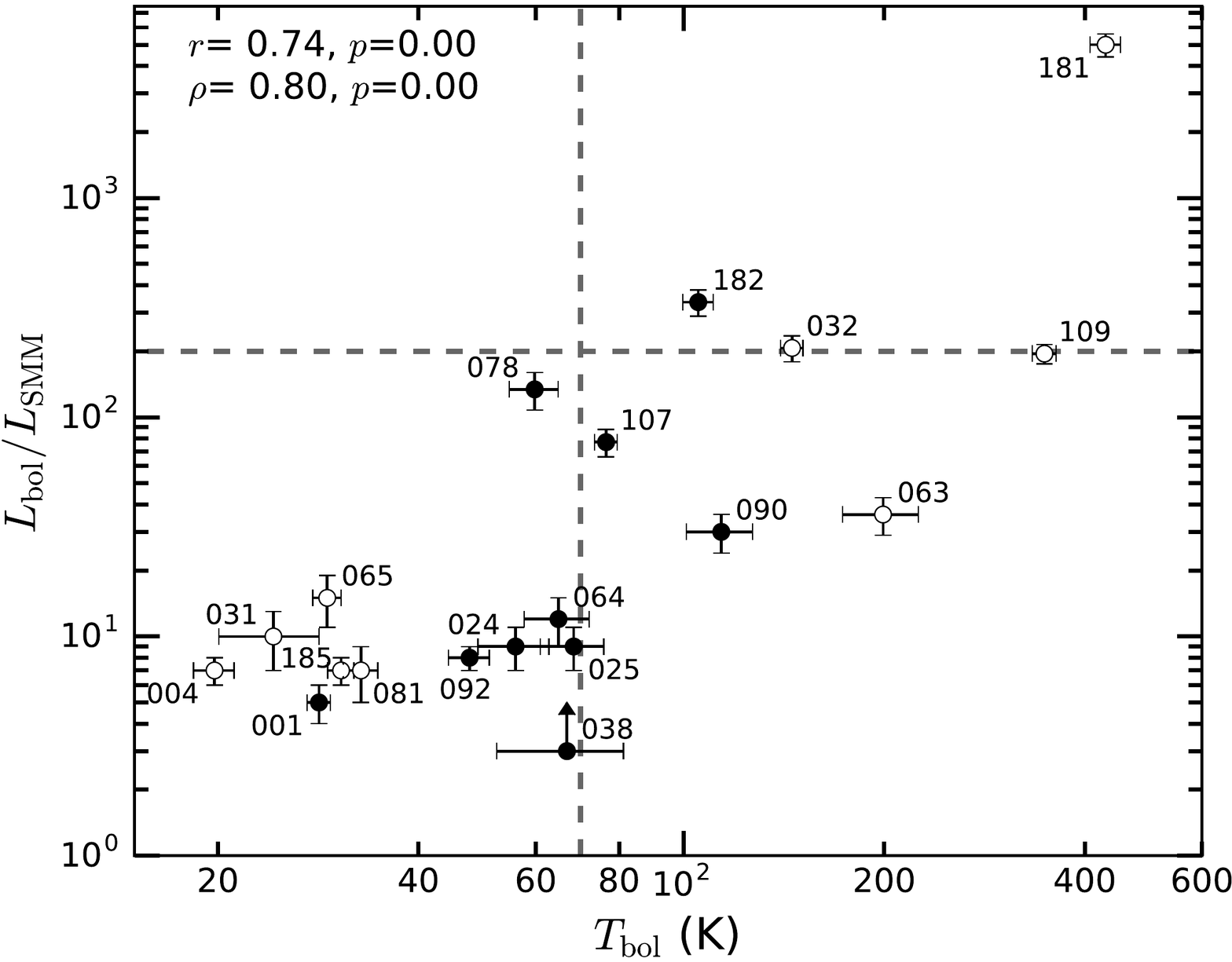}
\vspace{-90pt}
\caption{
Plot of observed bolometric temperature ($T_{\rm bol}$) versus the ratio of bolometric to submillimeter luminosity ($L_{\rm bol}/L_{\rm smm}$).
The black points indicate the sources that were fitted in the $T_{\rm bol}-\theta_{\rm open}$ plane ({\it i.e.} Figures \ref{fig:wid} and \ref{fig:wid2})
and the white points indicate those that were not.
The Pearson ($r$) and Spearman ($\rho$) correlation coefficients and their significance are displayed in the top left corner and are calculated
without the data point of DCE 038.
The dashed lines show the boundary of the Class 0 and Class I for each evolutionary indicator (\citealp[$T_{\rm bol}=70K$:][]{ch95}, \citealp[$L_{\rm bol}/L_{\rm smm}$=200:][]{an93}).}
\label{fig:TbolLL}
\end{figure}

\subsection{Outflow opening angles versus $L_{\rm bol}/L_{\rm smm}$ and other parameters from SED}
\label{sec:LL}
To study what physical conditions may affect the outflow opening angle, we compare the latter to four parameters from \citet{du08}: the internal luminosity ($L_{\rm int}$), the bolometric luminosity ($L_{\rm bol}$), the ratio of bolometric to submillimeter luminosity ($L_{\rm bol}/L_{\rm smm}$), and the submillimeter luminosity ($L_{\rm smm}$).
We exclude DCE 185 and 004 because their opening angles are considered as upper limits (see Sections \ref{sec:res} and \ref{sec:opeTbol}).
We evaluate the significance of correlation between the outflow opening angle and these parameters using Pearson's $r$ correlation test and Spearman's $\rho$ rank correlation test \citep{co99}.
Based on the correlation tests, we find no significant correlations between $\theta_{\rm open}$ and these parameters, except for $L_{\rm bol}/L_{\rm smm}$.
We find a correlation coefficient $r=0.55$ for $\theta_{\rm open}-L_{\rm bol}/L_{\rm smm}$, suggesting a probability (p-value) of 10\% for it being an uncorrelated/random distribution.
Since $L_{\rm bol}/L_{\rm smm}$ is considered as an evolutionary indicator \citep{an93,yo05}, we conclude that the opening angle is most likely sensitive to the evolutionary status but not to other physical conditions such as envelope mass traced by $L_{\rm smm}$.

\begin{figure}
\vspace{-70pt}
\includegraphics[scale=.42]{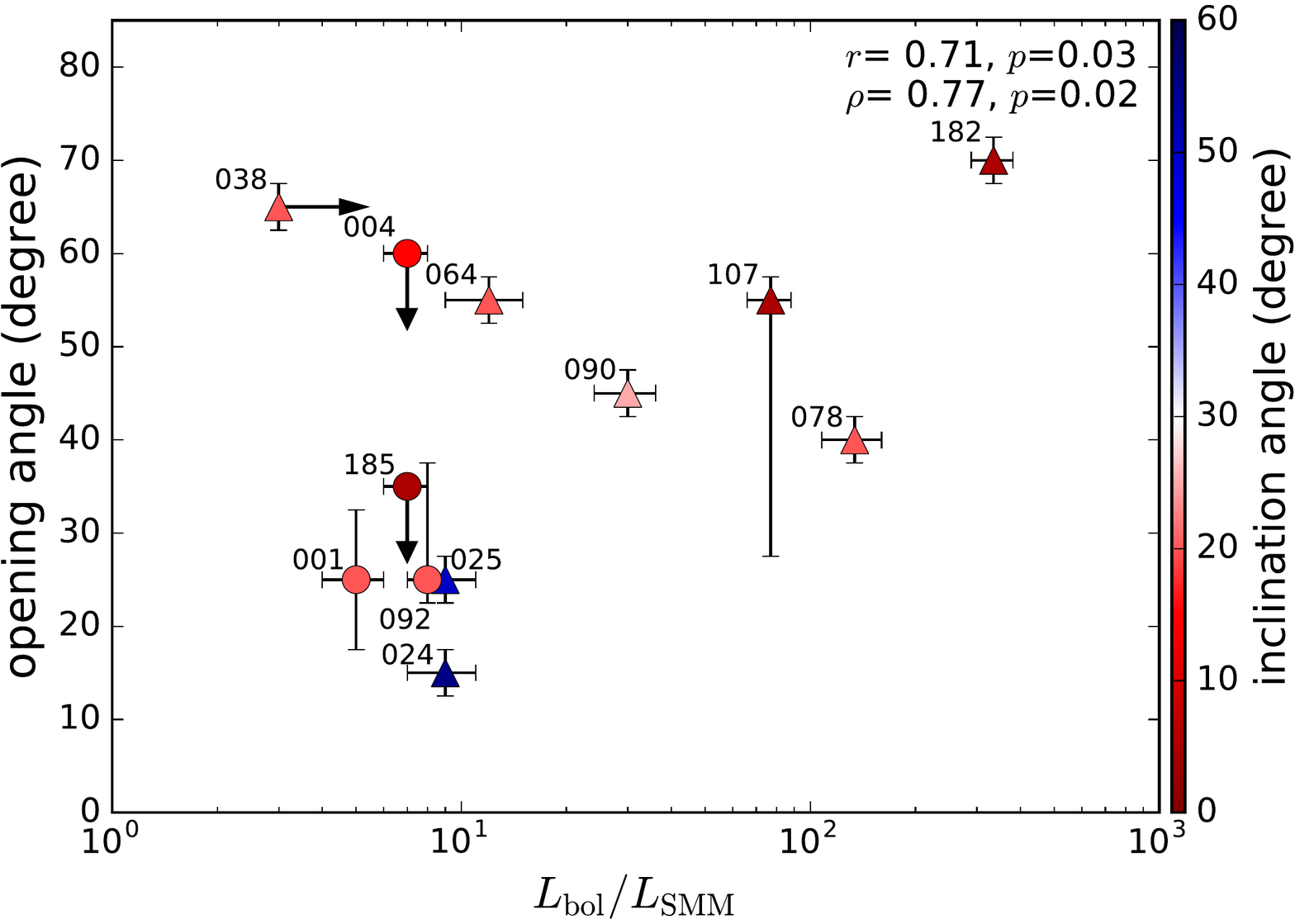}
\vspace{-100pt}
\caption{
Plot of outflow opening angles versus the ratio of bolometric to submillimeter luminosity ($L_{\rm bol}/L_{\rm smm}$).
The Pearson ($r$) and Spearman ($\rho$) correlation coefficients and their significance are displayed in the top right corner.}
\label{fig:par}
\end{figure}

\citet{yo05} suggest that $L_{\rm bol}/L_{\rm smm}$ can better reflect the evolutionary status than $T_{\rm bol}$ because, in their evolutionary model,
$L_{\rm bol}/L_{\rm smm}$ is sensitive to the fraction of mass accreted onto the central object but is less affected by the initial core mass.
Figure \ref{fig:TbolLL} shows our LLO sample in a $L_{\rm bol}/L_{\rm smm}$ versus $T_{\rm bol}$ diagram, along with the boundaries traditionally used to separate the Class 0 and I phases, either in bolometric temperature \citep[$T_{\rm bol}=70K$][]{ch95} or in bolometric-to-submm luminosity ratio \citep[$L_{\rm bol}/L_{\rm smm}$=200][]{an93}. The classification of our LLO sample into Class 0 and I objects depends somewhat on the chosen axis. Three objects out of 19 fall into the Class 0 category according to their $L_{\rm bol}/L_{\rm smm}$ ratio but into the Class I category according to their bolometric temperature. The classification of the rest of the sample is the same with both criteria.
Figure \ref{fig:TbolLL} shows a correlation between $T_{\rm bol}$ and $L_{\rm bol}/L_{\rm smm}$ toward the 19 LLOs ($r = 0.74$ and $\rho=0.72$ with both p-values $=0$), though the correlation is less significant when considering only the ten targets used to derive the $T_{\rm bol}-\theta_{\rm open}$ relation.
DCE 038 seems to be an outlier in both Figures \ref{fig:TbolLL} and \ref{fig:par} with an extremely low $L_{\rm bol}/L_{\rm smm}$ ratio (3 $\pm$ 1). This source has a puzzling jump in SED with 3.2 $\pm$ 0.5 Jy at 350 $\mu$m and 21.5 $\pm$ 16.1 Jy at 450 $\mu$m (1.8 $\pm$ 0.4 Jy at 850 $\mu$m), which results in the low $L_{\rm bol}/L_{\rm smm}$ \citep{du08} when the submillimeter luminosity is defined as the luminosity at $\lambda>350~\mu$m \citep{an93}. $L_{\rm bol}/L_{\rm smm}$ would become 10 $\pm$ 1 if we ignore the flux density at 450 $\mu$m or 6 $\pm$ 3 if we replace the observed flux density at 350 $\mu$m by a linear interpolation between 70 and 450 $\mu$m. This suggests that the observed $L_{\rm bol}/L_{\rm smm}$ is underestimated. Therefore, we remove DCE 038 and, in turn, find better correlation coefficients of $r = 0.74$ and $\rho=0.80$ between $T_{\rm bol}$ and $L_{\rm bol}/L_{\rm smm}$ (Figure \ref{fig:TbolLL}).
Furthermore, comparing to the outflow opening angle, we obtain correlation coefficients of $r = 0.71$ and $\rho=0.77$ (p-values of 0.03 and 0.02, respectively) for $L_{\rm bol}/L_{\rm smm}-\theta_{\rm open}$ (Figure \ref{fig:par}) and $r = 0.67$ and $\rho=0.70$ (p-values of 0.05 and 0.04, respectively) for $T_{\rm bol}-\theta_{\rm open}$.
Therefore, we suggest that the distribution of opening angles is better correlated with evolutionary status using $L_{\rm bol}/L_{\rm smm}$ as an age indicator than using $T_{\rm bol}$.
This result implies that the outflow opening angle likely reflects the evolutionary status well and could be considered as a good evolutionary indicator.

\section{SUMMARY}
\label{sec:summary}
We conducted an outflow/jet survey of 20 Low Luminosity Objects (LLOs) using CFHT in Ks-band continuum emission and H$_2$ line emission.
From the Ks-band observations, we identify a background galaxy and eliminate it from our sample.
We detect outflow cavities in eight sources out of 19 LLOs on the basis of the Ks-band continuum observations.
Among the remaining 11 LLOs, the archival \textit{Spitzer} IRAC1 data reveals four outflow-driving sources.
We derive the outflow opening angles and the inclination angles of these 12 LLOs by fitting the observed images with the radiative transfer models of \citet{wh03a}.
The H$_2$ observations reveal only one LLO associated with a prominent H$_2$ jet. 
Three other LLOs out of the remaining 18 have marginal detections of an H$_2$ jet.
Our main results are the following:

\begin{enumerate}
\item 
As we detect only one to four H$_2$ jets out of 19 LLOs, we suggest that most LLOs do not generate strong jets or outflows.
This indirectly supports the idea that these LLOs are likely at a quiescent accretion phase.
\item 
Our LLOs follow a trend similar to the one found by AS06 in the plot of bolometric temperature versus opening angle ($T_{\rm bol}-\theta_{\rm open}$), and are broadly consistent with the correlation reported by VLT14, after correction for the inclination.
Instead of a single power-law describing the distribution, we find a larger index for the sources at an early evolutionary stage ($T_{\rm bol}$ $\lesssim$ 120 K), in agreement with VLT14.
We conclude that the outflow opening angle may better trace the evolutionary stage than $T_{\rm bol}$ for $T_{\rm bol} < 120$~K.
\item
The LLOs located above the best-fit power-law index have relatively small inclination angles and those located below have large inclination angles. 
This suggests that the dispersion in the $T_{\rm bol}-\theta_{\rm open}$ plot may be in part due to inclination effects on $T_{\rm bol}$.
\item 
Using the outflow opening angle as an evolutionary indicator, we suggest that
at least 40\% of the outflow-driving LLOs in our sample are young 
Class 0 objects.
\item
Out of the 12 targets with infrared outflow detections, ten are better fitted with a curved cavity than a streamline cavity, suggesting that the outflow cavities are not carved by precessing jets.

\end{enumerate}

\acknowledgments
The authors thank Dr. Masaaki Hiramatsu for fruitful discussions and a comparison to his SMA outflow survey of LLOs in Perseus.
We are grateful to Dr. Chi-Hung Yan for his assistance with the CFHT data reduction.
We would like to thank Prof.\ H\'{e}ctor Arce, Dr.\ Ian Stephens, and Dr.\ Thomas Robitaille for providing valuable discussions. 
The authors acknowledge the staff at CFHT for assistance with operations.
The authors thank the referee for the insightful comments that improved this paper.
T.H.H. and S.P.L. are thankful for the support of the Ministry of Science and Technology (MoST) of Taiwan through Grants NSC 98-2112- M-007-007-MY3, NSC 101-2119-M-007-004, MoST 102-2119-M-007-004-MY3 and MoST 105-2119-M-007-024.

\end{document}